\documentclass[runningheads]{llncs}
\usepackage[T1]{fontenc}
\usepackage{graphicx}
\usepackage{subfiles}

\usepackage{amsmath}
\usepackage{siunitx}
\usepackage{upgreek}
\usepackage{chemmacros}

\usepackage[super]{nth}

\begin{document}
\title{Measuring the limit of perception of bond stiffness of interactive molecules in VR via a gamified psychophysics experiment}
\titlerunning{Measuring the limit of perception of bond stiffness}
%
\author{
Rhoslyn Roebuck Williams\inst{1}\orcidID{0000-0003-2535-7180} \and
Jonathan Barnoud\inst{1}\orcidID{0000-0003-0343-7796} \and
Luis Toledo\inst{1}\orcidID{0009-0001-4683-9835} \and
Till Holzapfel\inst{1,2} \and
David R. Glowacki\inst{1}\orcidID{0000-0002-9608-3845}
}
\authorrunning{R. Roebuck Williams et al.}
%
\institute{CiTIUS$\sim$Centro Singular de Investigaci\'on en Tecnolox\'ias Intelixentes, Santiago de Compostela, Spain,
\email{rhoslyn.roebuckw@usc.es} \\
\url{https://www.intangiblerealitieslab.org/} \and
Neuroinformatics Group, University of Osnabrueck, Germany}
\maketitle              
\begin{abstract}
Molecular dynamics (MD) simulations provide crucial insight into molecular interactions and biomolecular function. With interactive MD simulations in VR (iMD-VR), chemists can now interact with these molecular simulations in real-time. Our sense of touch is essential for exploring the properties of physical objects, but recreating this sensory experience for virtual objects poses challenges. Furthermore, employing haptics in the context of molecular simulation is especially difficult since \textit{we do not know what molecules actually feel like}. In this paper, we build upon previous work that demonstrated how VR-users can distinguish properties of molecules without haptic feedback. We present the results of a gamified two-alternative forced choice (2AFC) psychophysics user study in which we quantify the threshold at which iMD-VR users can differentiate the stiffness of molecular bonds. Our preliminary analysis suggests that participants can sense differences between buckminsterfullerene molecules with different bond stiffness parameters and that this limit may fall within the chemically relevant range. Our results highlight how iMD-VR may facilitate a more embodied way of exploring complex and dynamic molecular systems, enabling chemists to sense the properties of molecules purely by interacting with them in VR. 

\end{abstract}

\section{Introduction}
Scientific concepts are often abstract and exist outside the realm of our embodied knowledge and everyday sensory experience. This is especially true for chemistry, where the objects of study (electrons, atoms, molecules) lie tantalisingly out of reach of our senses. As such, tools for `scaling up' the nanoscale have been developed that reveal this previously-hidden molecular world. Molecular dynamics (MD) simulations, in which the nanoscale movements of molecules are modelled using classical mechanics, provide important insight into molecular interactions \& biomolecular function, and are now routinely used in areas such as computational drug design \cite{sabe_current_2021}. 
Advancements in computer power and increasing availability of virtual reality (VR) have made possible \textit{immersive} visualisation of these molecular systems. These technologies facilitate more efficient and natural exploration of complex and dynamic molecular structures, thus enabling the researcher to gain chemical insight in a more intuitive manner.

As well as immersive visualisation, we are now able to \textit{interact} with these molecular simulations in real time in VR. Using interactive molecular dynamics in virtual reality (iMD-VR), VR users can apply forces to molecular systems using natural hand movements as the simulations are calculated in real time \cite{oconnor_interactive_2019}. In this way, the chemist is able to enter the nanoscale world to interact and explore molecular structures in a way that is more akin to how we would handle and explore physical objects. iMD-VR has been used for several applications, including simulating ligand docking \cite{mishra_molecular_2024,deeks_interactive_covid_2020}, teaching enzyme catalysis \cite{bennie_teaching_2019} and chemical reactivity \cite{seritan_interachem_2021}, and investigating reactive chemical systems \cite{amabilino_training_2020}. Shannon et al. \cite{shannon_exploring_2021} successfully employed gamification within iMD-VR to explore reaction pathways, finding that citizen scientists generated almost all the important pathways known in the existing literature.

Our sense of touch is important for understanding the properties of physical objects. Historically, most interactive molecular simulations have provided haptic feedback, typically employing desktop-mounted pointers and arms. As the user moves the device, it provides a resistance that is related to the forces acting within the molecular system. However, haptic technologies still face many limitations, including high costs, low availability \& accessibility, and, more fundamentally, their current inability to stimulate the full breadth of physical touch. Furthermore, a problem arises specifically in the context of molecular simulations: there is no physical analogue from which to deduce and compare the haptic properties of molecules. In other words, since we cannot reach out and touch a molecule, \textit{we do not know what molecules actually feel like}. Therefore, each mapping from a particular molecular property (e.g. the potential energy) onto some dimension of haptic feedback (e.g., controller vibrations) is only one of the many possible ways to represent what it feels like to touch a molecule. The presented work uses ideas from the human-computer interaction literature to illuminate an alternative way of sensing the properties of molecules in VR which does not require haptic feedback.

L\'ecuyer et al. \cite{lecuyer_pseudo-haptic_2000} coined the term `pseudo-haptic feedback' to describe how an interactive digital environment can convey information about the mechanical properties of virtual objects through visual feedback. Pseudo-haptics has since been investigated for conveying a range of properties, including force feedback \cite{onishi_bouncyscreen_2021}, stiffness \cite{weiss_using_2023}, and weight \cite{kim_effect_2022} (see this review paper \cite{ujitoko_survey_2021}). 

This idea was first extended to molecular simulations by Roebuck Williams et al. \cite{roebuck_williams_subtle_2020}, who found that VR users could discriminate between identically-shaped molecules of different rigidity. Furthermore, the authors found that participants who were \textit{interacting} with the molecules were better at distinguishing these properties than those who were solely observing whilst cohabiting the same virtual space. In this paper we present the results of a pilot study that extends this initial work with the aim of quantifying this limit of perception through a two-alternative forced choice (2AFC) psychophysics experiment. The objectives of this study were to (a) approximate the limit of perception of bond stiffness of buckminsterfullerene molecules, and (b) investigate whether this limit of perception is different when using VR controllers (using the trigger button to interact) and hand tracking (using an index pinch to interact). Furthermore, the presented work implements a gamified experimental design, inspired by Shannon et al.'s \cite{shannon_exploring_2021} previous use of gamification within iMD-VR.

\section{Materials and methods}
This study was performed using \textit{Subtle Game}, a single-player iMD-VR game that was developed using the NanoVer iMD-VR framework \cite{jamieson-binnie_narupa_2020} and is available open-source on GitHub (\url{https://github.com/IRL2/SubtleGame}). Subtle game facilitates three mini-games, which are explained to the player inside VR: (a) the \textit{trials task}, constituting the 2AFC psychophysics experiment; (b) a \textit{nanotube task}, where the player pulls a methane molecule through the centre of a carbon nanotube; and (c) a \textit{knot-tying task}, where the player ties a trefoil knot in the 17-alanine polypeptide. For this study, the nanotube and knot-tying tasks provided participants with training for interacting with the molecular simulations and were inserted between rounds of psychophysics trials. Data from these tasks are being processed as part of other ongoing research and their analysis are beyond the scope of this paper.

\begin{figure}
    \centering
    \includegraphics[width=\textwidth]{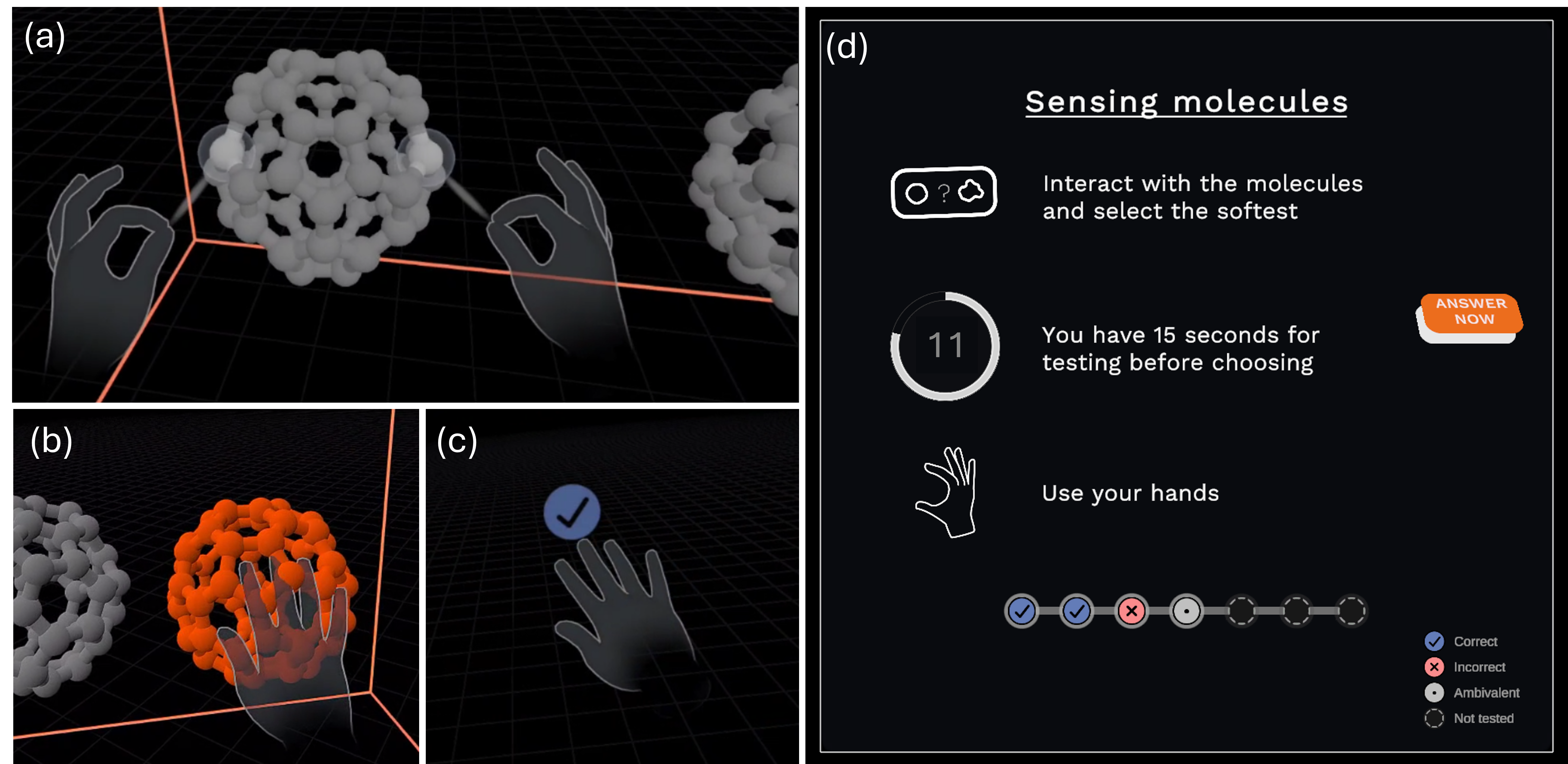}
    \caption{Showing first-person perspective screenshots of a player using their hands to perform the 2AFC psychophysics experiment: (a) a player interacting with the molecules, (b) selecting their answer, (c) feedback on whether the answer was correct, and (d) a UI panel with additional information that is visible during the task.}
    \label{fig:psychophysics-trials-screenshots}
\end{figure}

The trials task (Fig. \ref{fig:psychophysics-trials-screenshots}) consists of rounds of psychophysics trials in which the player is given 15 seconds to interact with two buckyball molecules. Once the time is up, the player must answer which molecule was the easiest to deform (i.e. softest), after which they are informed whether their answer was correct. Participants were told that sometimes the two buckyballs would be identical, in which case they should choose one at random. In each trial, one of the buckyballs was defined using experimentally-derived parameters, and the stiffness of the other buckyball was modified by scaling the \textit{angle force constants} (further information below). We chose buckyballs for two reasons: (1) buckyballs are made up of only one type of atom (carbon), so a consistent change in property could be achieved by scaling the force constants by a single value, and (2) this change in property can be described simply as a change in rigidity.

\subsubsection{Preparing the simulations.} The buckyball simulations used in Subtle Game are similar to the buckyball simulations detailed in previous publications \cite{oconnor_sampling_2018,oconnor_interactive_2019}. The simulation file specifies: (a) anharmonic bond and angle forces defining the stiffness of bonds and between neighbouring bonds; (b) a custom nonbonded force describing electrostatic and van der Waals interactions between nonbonded atoms; and (c) other parameters such as temperature (\SI{300}{\kelvin}), integrator (Langevin), friction coefficient (\SI{1}{\per\pico\second}), and timestep (\SI{1}{\femto\second}). To alter the rigidity of a buckyball, we scaled the experimentally-derived angle force constant ($k = \SI{458}{\kilo\joule\per\mol\per\radian\squared}$), keeping all other parameters identical. Scaling this value modifies how easily the buckyball could be deformed out of the equilibrium (soccer ball) shape. Scaling by a value $<1$ results in a softer, more easily deformed buckyball and scaling by a value $>1$ results in a harder, more rigid buckyball. The nanotube + methane and the 17-alanine polypeptide simulations are also detailed in previous publications \cite{oconnor_sampling_2018,oconnor_interactive_2019}, with only minor modifications to the simulation parameters. All simulation files are provided in the Subtle Game git repo.

We conducted a short experiment to estimate the upper and lower detection limits to select the scaling values used in the pilot study. We initially estimated that a scaling of $0.4 x$ and $1.6 x$, where $x$ is the experimentally-derived angle force constant value, were large enough for most people to sense a difference. Four participants took part in an ascending and descending method of limits experiment for a softer comparison stimulus (0.4$x$--1$x$) and a harder comparison stimulus (1$x$--1.6$x$), using increments of $0.1$ since the force constants could not be scaled continuously. The tests were performed using hand tracking and controllers, and the detection points were averaged across participants and interaction modes. These calculations yielded a lower bound of $0.75 x$ ($\mathrm{SD}=0.20$) and an upper bound of $1.30 x$ ($\mathrm{SD}=0.20$), rounded to 2 d.p. to the nearest 0 or 5. Based on these results, we decided to use the following scaling values in the pilot study: $0.625 x$, $0.75 x$, $0.875 x$, $1 x$, $1.125 x$, $1.25 x$, and $1.375 x$.

\subsection{Study design}
\subsubsection{Participants.} Thirteen participants (6 female, 5 male, 2 non-binary/third gender) were recruited via convenience sampling at the University of Bologna, Italy. The mean age was 28 years ($\mathrm{SD}=7$) and there was a varied experience of VR, with 4 participants reporting that this was their first time using VR, and 3 participants who used VR 50$+$ times. Most (11 participants) had no higher-level qualification in molecular modelling or related field. 

\subsubsection{Experimental design.} This study was approved by the Bioethics Committee at the University of Santiago de Compostela, Spain (USC 51/2022). A within-subjects design approach was employed in which each participant performed two `rounds' of Subtle Game using one interaction mode, and then performed two more rounds with the other interaction mode. Each game round consisted of performing each of the three tasks once in the following order: (1) the nanotube task, (2) 21 $\times$ psychophysics trials (3 repeats for each scaling value), (3) the knot-tying task. This was a total of $84$ trials per participant. The order of interaction mode was randomised, with 8 participants using hand tracking first and controllers second, and vice versa for the other 5 participants. 

We used a laptop running Windows 11 with an AMD Ryzen 9 6900HX processor and a GeForce RTX 3070 Ti GPU (the same laptop was used for all participants). The VR game was run on the laptop through the Unity Editor, with each participant using a Meta Quest 2 HMD connected via a link cable to the laptop. The NanoVer server and a python script running the game logic were also running locally on the laptop. 

Each experiment took approximately 1 hour and started with preparation, where participants were given an informed consent form and information sheet, with time to ask questions to the experimenter. Participants then began the VR portion of the study: (1) performing two rounds of the game using the one interaction mode; (2) taking a short break, which ended when the participant was ready to start the next part ($\sim$5 minutes on average); and (3) re-entering VR and repeating the game using their second interaction mode. Afterwards, they filled out an online questionnaire (13 responses) and took part in a semi-structured interview (10 responses, due to time constraints). All participants completed the full study, however, the data for 3 participants were only partially recorded  (\SI{75}{\percent}, \SI{50}{\percent}, and \SI{50}{\percent}) due to technical issues (see the supplementary information for the full dataset).

\section{Results and discussion}
\begin{figure}[t!]
    \centering
    \includegraphics[trim={0.5cm 3cm 3cm 3cm},clip,width=\textwidth]{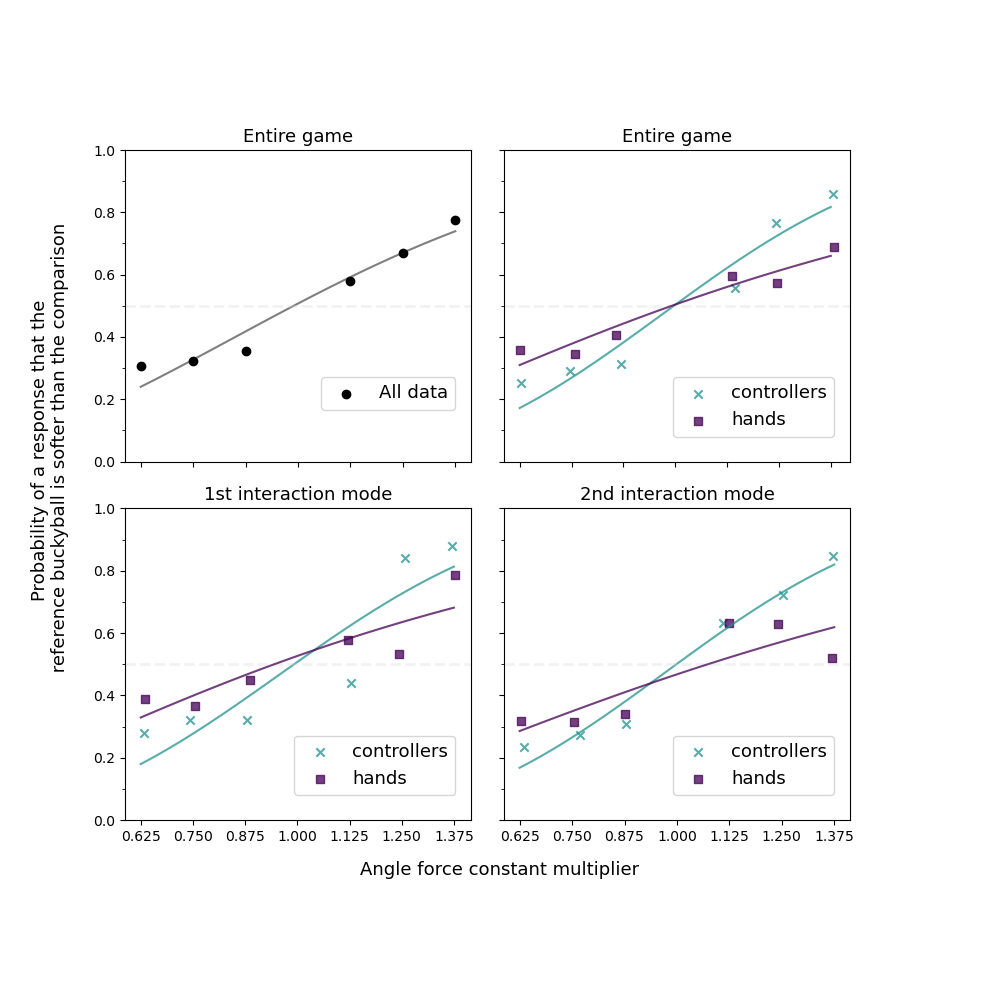}
    \caption{Showing the results of the 2AFC experiment for (a) all data across the entire game (top left), (b) the data for hand tracking vs. controllers across the entire game (top right), (c) the data for hand tracking vs. controllers when it was participants' first interaction mode (bottom left), the data for hand tracking vs. controllers when it was participants' second interaction mode. The curves were fitted to a Weibull cumulative distribution function using \texttt{scipy.optimise.curve\_fit} in a Python script.}
    \label{fig:psychometric_curves}
\end{figure}

The psychophysics trials were processed by removing responses for trials in which the molecules were identical. We then calculated the probability of a response that the reference buckyball was \textit{softer} than the comparison buckyball. We plotted these probabilities and fitted a Weibull cumulative distribution function (commonly used in 2AFC experiments \cite{mortensen_additive_2002}) of the form given in equation \eqref{eq:weibull_cum_dist}. This process was repeated using several subsets of the data (Fig. \ref{fig:psychometric_curves}).
\begin{equation}
    \text{w}'' (x) = 1 - e^{-(x / \lambda) ^\kappa}
\label{eq:weibull_cum_dist}
\end{equation}

We used a threshold of \SI{25}{\percent} to calculate the lower JND (JND$_l$) and \SI{75}{\percent} for the upper (JND$_u$) from the fitted curves (Table \ref{tab:JND_values}). The values across the entire dataset were JND$_l = 0.64$ and JND$_u = 1.4$. However, when separating the data for controllers and hands, the data show that participants were better able to distinguish differences with controllers (JND$_l = 0.73$, JND$_u = 1.28$) than hand tracking (JND$_l = 0.52$, JND$_u = 1.65$). The value of the \ch{C-C-C} angle force constant of the reference buckyball is \SI{458}{\kilo\joule\per\mol\per\radian\squared}, thus, the controller JND values correspond to participants being able to distinguish differences of $-124$ and $+$\SI{128}{\kilo\joule\per\mol\per\radian\squared}. These values are on the order of differences between typical angle force constants, e.g. those used to define the 17-alanine simulation used in the knot-tying task. However, this could be in part due to the highly structured shape of the buckyballs and further work is needed to assess how these JND values change for different molecules. 

\renewcommand{\arraystretch}{1.25}
\setlength{\tabcolsep}{6pt}
\begin{table}[t!]
\centering
\begin{tabular}{c | c c c | c c c | c c c}
 & \multicolumn{3}{c|}{\textbf{entire game}} & \multicolumn{3}{c|}{\textbf{\nth{1} mode}} & \multicolumn{3}{c}{\textbf{\nth{2} mode}} \\
 & A & H & C & A & H & C & A & H & C \\
 \hline\hline
JND$_l$ & 0.64 & 0.52 & 0.73 & 0.59 & 0.49 & 0.72 & 0.68 & 0.56 & 0.73 \\
JND$_u$ & 1.40 & 1.65 & 1.28 & 1.41 & 1.59 & 1.28 & 1.38 & 1.80 & 1.28 \\
\end{tabular}
\caption{Showing the lower (JND$_l$) and upper (JND$_u$) just-noticeable-difference values for the buckyball angle force constants, which were obtained from the fitted Weibull cumulative distribution curves (see Fig. \ref{fig:psychometric_curves}) at probability values of 0.25 and 0.75 respectively and are given to 2 decimal places. The values were calculated for responses across (a) the entire game, (b) the data from participants' first interaction mode, and (c) the data from participants' second interaction mode. Within each of these subsets, calculations were performed on all the data (`A'), only the hand tracking data (`H'), and only the controller data (`C'). } 
\label{tab:JND_values}
\end{table}

As well as better JNDs, the controllers data are more consistent than the hand tracking data. For example, the values are almost identical when participants used controllers as their first (JND$_u = 0.72$, JND$_l = 1.28$) and second interaction mode (JND$_u = 0.73$, JND$_l = 1.28$), suggesting that there is not a large learning effect when using controllers. This contrasts with the hands, where the data vary with order of interaction mode. For example, participants using hands as their first interaction mode were more effective at sensing differences when the comparison buckyball was harder, but worse with the softer comparison, however, this trend is reversed for participants using hands as their second interaction mode. We observed that many participants struggled in general using the hand tracking, in part because many attempted to use large hand movements, which reduces the tracking accuracy of the hands---a limitation of the hand tracking of the HMD---thus making the task more difficult. 

In the post-study questionnaire, participants were asked about their preferences for controllers and hands for each task. The results showed that more participants preferred the controllers overall ($n=8$) than the hands ($n=4$), with 1 participant having no overall preference. This trend was more evident for the psychophysics trials, with 10 participants preferring controllers compared to only 1 participant preferring hands. There was also a tendency for participants to have an overall preference for their second interaction mode, with 9 participants choosing their second interaction mode---controllers ($n=6$) and hands ($n=3$)---compared to 3 who chose their first interaction mode---controllers ($n=2$) and hands ($n=1$). However, this may be because of a general preference for controllers combined with the fact that more participants had controllers as their second interaction mode ($n=8$). 

\subsection{Conclusions and future work}
In a previous study, Roebuck Williams et al. \cite{roebuck_williams_subtle_2020} demonstrated that VR-users could distinguish the stiffness of bonds of interactive molecular simulations in VR, showing that participants could distinguish between buckyballs for which the angle force constants had been scaled by $\frac{1}{10}$, $1$, and $10$. Employing methods from the psychophysics literature, we built upon this initial study by using a 2AFC experimental design to quantify the limit of this perception. Our initial results suggest that this limit may be more fine-grained than previously demonstrated, on the order of $0.73 x$ and $1.28 x$, which is within the chemically relevant range. These preliminary results illustrate how iMD-VR might facilitate a more intuitive mode of interaction with molecular systems, which can enable a researcher to sense molecular properties purely through interacting with them in VR. The data also indicate that this perception may be best when using VR controllers compared to using hands via hand tracking. 

Furthermore, whereas in the previous study participant responses were collected post-VR, Subtle Game employs a gamification approach in which tasks are explained to the player in-game and responses are collected on-the-fly, i.e. whilst the player is in VR. This design approach is a step towards facilitating unsupervised, citizen science data collection. We are currently working on mounting the behind-the-scenes game logic onto a cloud server, which would allow any citizen scientist to download Subtle Game onto their VR headset and play for a flexible period of time. This larger-scale data collection would allow us to investigate the limit of perception for a range of different molecules, and further investigate how interaction mode may modulate this perception.

\subsubsection{Acknowledgements} We wish to thank Professor Emanuele Paci and his research group at the University of Bologna for hosting us during data collection, and Sila Sobrado for her support and assistance throughout this project. This project was supported by the Xunta de Galicia (Centro de investigación de Galicia accreditation 2019–2022, ED431G-2019/04) and the European Union (European Regional Development Fund—ERDF), as well as by the European Research Council under the European Union’s Horizon 2020 research and innovation program through consolidator grant NANOVR 866559.



\bibliographystyle{splncs04}
\bibliography{bibliography}

\end{document}